\documentclass{ieeeaccess_modified}
\usepackage{cite}
\usepackage{amsmath,amssymb,amsfonts}
\usepackage{algorithmic}
\usepackage{graphicx}
\usepackage{textcomp}
\usepackage{upgreek}
\usepackage{soul}

\newlength{\xfigwd}
\def\BibTeX{{\rm B\kern-.05em{\sc i\kern-.025em b}\kern-.08em
    T\kern-.1667em\lower.7ex\hbox{E}\kern-.125emX}}
    
\begin{document}
\history{}
\doi{}

\title{Experimentally verified, fast analytic and numerical design of superconducting resonators in flip-chip architectures}

\author{\uppercase{Hang-Xi Li}\authorrefmark{1}, \uppercase{Daryoush Shiri}\authorrefmark{1}, 
\uppercase{Sandoko Kosen}\authorrefmark{1}, \uppercase{Marcus Rommel}\authorrefmark{1}, \uppercase{Lert Chayanun}\authorrefmark{1}, \uppercase{Andreas Nylander}\authorrefmark{1}, \uppercase{Robert Rehammar}\authorrefmark{1}, \uppercase{Giovanna Tancredi}\authorrefmark{1}, \uppercase{Marco Caputo}\authorrefmark{2}, \uppercase{Kestutis Grigoras}\authorrefmark{2}, \uppercase{Leif Grönberg}\authorrefmark{2}, \uppercase{Joonas Govenius}\authorrefmark{2}, and \uppercase{Jonas Bylander}\authorrefmark{1}
}

\address[1]{Department of Microtechnology and Nanoscience, Chalmers University of Technology, 412 96 Gothenburg, Sweden}
\address[2]{VTT Technical Research Centre of Finland Ltd., QTF Centre of Excellence, FI-02044 VTT Espoo, Finland}

\tfootnote{This work was funded by the Knut and Alice Wallenberg (KAW) Foundation through the Wallenberg Center for Quantum Technology (WACQT) and by the EU Flagship on Quantum Technology H2020-FETFLAG-2018-03 project 820363 OpenSuperQ and HORIZON-CL4-2022-QUANTUM-01-SGA project 101113946 OpenSuperQPlus100. We acknowledge the use of support and resources from Myfab Chalmers, and Chalmers Center for Computational Science and Engineering (C3SE, partially funded by the Swedish Research Council through Grant Agreement No. 2018-05973).}

\markboth
{Li \headeretal: Experimentally verified, fast analytic and numerical design of superconducting resonators in flip-chip architectures}
{Li \headeretal: Experimentally verified, fast analytic and numerical design of superconducting resonators in flip-chip architectures}

\corresp{Corresponding author: Hang-Xi Li (email: hangxi@chalmers.se).}

\begin{abstract}
In superconducting quantum processors, the predictability of device parameters is of increasing importance as many labs scale up their systems to larger sizes in a 3D-integrated architecture. In particular, the properties of superconducting resonators must be controlled well to ensure high-fidelity multiplexed readout of qubits. Here we present a method, based on conformal mapping techniques, to predict a resonator's parameters directly from its 2D cross-section, without computationally heavy and time-consuming 3D simulation. We demonstrate the method's validity by comparing the calculated resonator frequency and coupling quality factor with those obtained through 3D finite-element-method simulation and by measurement of 15 resonators in a flip-chip-integrated architecture. We achieve a discrepancy of less than 2\% between designed and measured frequencies, for 6-GHz resonators. We also propose a design method that reduces the sensitivity of the resonant frequency to variations in the inter-chip spacing.
\end{abstract}

\begin{keywords}
conformal mapping, coplanar waveguide, flip chip, kinetic inductance, penetration depth, quantum processor, superconducting resonator
\end{keywords}

\titlepgskip=-15pt

\maketitle


\section{Introduction}
\IEEEPARstart{A}{s} the size of superconducting quantum processors grows beyond a small number of qubits, more advanced circuit-integration technology needs to be developed to accommodate the increasing input/output (I/O) wiring complexity. To this end, flip-chip integration is one choice that allows the routing of individual I/O lines to address qubits in the interior of the processor\cite{Rosenberg2017,Foxen2018,Arute2019}. Flip-chip integration technology has been demonstrated to be compatible with high-quality qubits\cite{Arute2019,Jurcevic2021,Kosen2021,Acharya2022,Zhang2022,Zhang2023,Kim2023}. Flip-chip modules comprise two or more dies that have been aligned and bonded together -- in the minimal version using a ``qubit" tier and a ``control" or ``interposer" tier, separating the fabrication of qubits and I/O circuitry onto different tiers. To achieve high-performance qubit control and readout, the measured resonant frequencies and other device parameters must agree with their design targets. The electromagnetic analytic and numerical design methodology also needs to be efficient, since 3D electromagnetic modeling of the full circuit is computationally intensive. A conventional simulation strategy is to sweep multiple design parameters of the resonator geometry to meet target $f_r$ and $Q_c$ values~\cite{Kosen2021}. While this is appropriate for small devices, scaling it beyond a few resonators becomes too time-consuming. 

In this paper, we present a design methodology for superconducting flip-chip-integrated quantum processor components. We apply it to the calculation of resonant frequencies ($f_r$) of coplanar-waveguide (CPW) resonators and their coupling quality factors to a readout feedline ($Q_c$). By analyzing only the 2D cross section of the CPW, we calculate these quantities using two methods: conformal mapping analysis and 2D numerical simulation. We compare the results to those obtained by 1000-times more resource demanding 3D simulation, and to experimental measurements of 15 aluminum resonators in a flip-chip quantum processor architecture. With the inclusion of the kinetic inductance contribution, we predict the resonant frequencies with accuracy better than 100~MHz.

The inter-chip spacing can be off-target, or may vary across a flip-chip module. 
For a resonator that faces a ground plane on the opposing chip, we propose a design method with a partial ground-plane cutout of the area directly facing the resonator, reducing the sensitivity of its resonant frequency to the inter-chip spacing. 

\section{2D cross-section calculations}
\label{section: 2D cross-section calculations}

\begin{figure*}[t]
    \centering
    \includegraphics[width=0.9\textwidth]{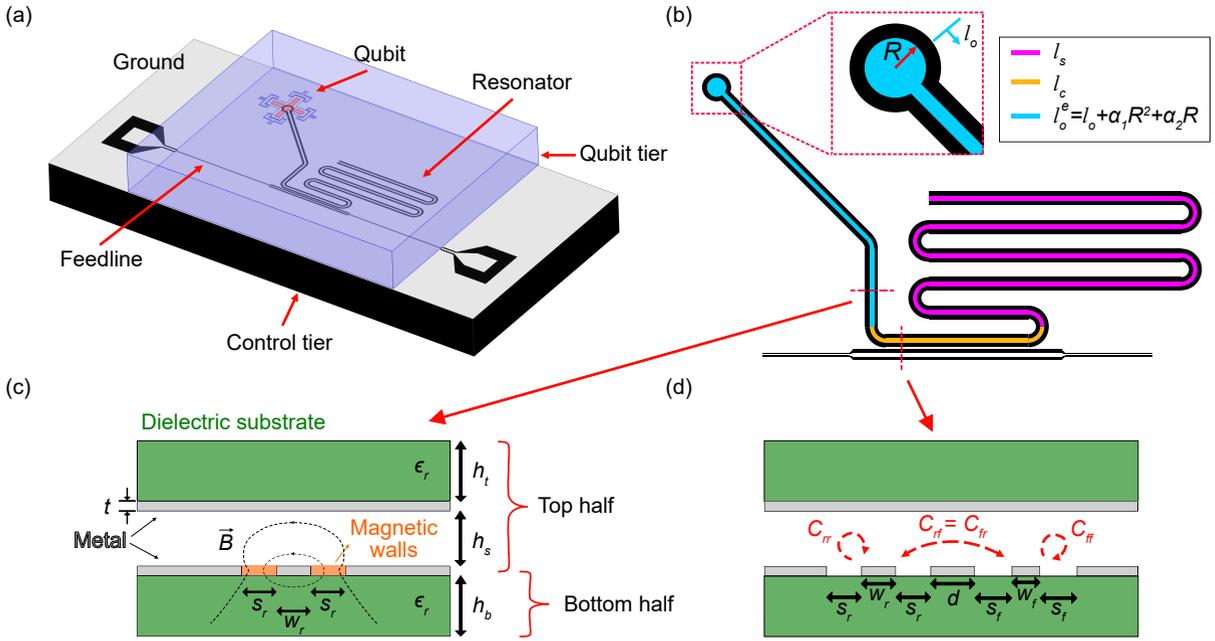}
    \caption{(a) False-color 3D model of a $\lambda/4$ resonator coupled to a feedline, on the control tier (bottom), facing a metal ground plane on the qubit tier (top) with a qubit located opposite from the open end of the resonator. (b) Top view of the resonator design. The resonator can be separated into three parts: effective open part $l_o^e$ (cyan), coupling part $l_c$ (yellow), and short part $l_s$ (pink). The effective open part contains both an open part $l_o$ and an effective length ($\alpha_1R^2+\alpha_2R$) due to the circular coupling structure. All the gap areas (black) of the resonator, including the circular coupling structure at the open end, have the same distance to the ground plane. (c) Cross-section of the resonator's CPW. The top and bottom halves of the cross-section, separated by the metal surface of the control tier (bottom), can be independently transformed into two parallel plates, using conformal mapping techniques, when the magnetic walls (orange) are placed on the surfaces of the CPW's gap area. (d) Cross-section of the resonator's coupling part with the feedline. The capacitance between the resonator's and the feedline's center conductors ($C_{rf}=C_{fr}$) and their self capcitances ($C_{rr},\,C_{ff}$) are simulated to obtain the coupling quality factor $Q_c$ to the feedline. }
    \label{fig:3D2DRes_topMetal}
\end{figure*}

In this section we show the calculation of $f_r$ and $Q_c$ from a CPW resonator's 2D cross-sections in flip-chip geometry with an inter-chip vacuum gap. 
Fig.~\ref{fig:3D2DRes_topMetal}(a) shows a 3D model of a typical quarter-wavelength ($\lambda/4$) resonator resting on the control tier, coupled to a feedline, also on the control tier, and to a qubit on the opposing qubit tier. We focus on the typical scenario in which the whole CPW resonator and feedline are facing a metal ground plane on the qubit tier~\cite{Rosenberg2017,Kosen2021}.

\subsection{Resonant frequency}
The resonant frequency of a $\lambda/4$ resonator is calculated by
\begin{equation}
    f_r = \frac{1}{4l_{tot}\sqrt{(L_l^g+L_l^k)\cdot C_l}},
    \label{eqn: fr_2D}
\end{equation}
where $l_{tot}$ is its total length including the effective lengths resulting from CPW discontinuities at both ends. $L_l^g$ and $L_l^k$ are the geometric and kinetic inductances per unit length, respectively, and $C_l$ is the capacitance per unit length. 

We can obtain $L_l^g$ and $C_l$ by using either conformal mapping techniques ($L_l^{g,\text{conf}}$, $C_l^{\text{conf}}$)~\cite{schinzinger2003} or 2D finite-element-method (FEM) simulation ($L_l^{g,\text{sim2D}}$, $C_l^{\text{sim2D}}$)~\cite{ansys2D} of the resonator's CPW cross-section as illustrated in Fig.~\ref{fig:3D2DRes_topMetal}(c). Here $h_b=h_t$ represent the substrate thicknesses of the control and qubit tiers,  $h_s$ is the inter-chip spacing, and $w_r$ and $s_r$ are the width and gap of the CPW line, respectively. We will examine the accuracies of calculations of $f_r$ using both 2D approaches.

To implement the conformal mapping techniques, we assume that the metal thin films of the resonator's central conductor and of the ground planes on both tiers have infinite conductivity, and that the CPW cross-section satisfies the quasi-TEM approximation. In order to simplify the conformal transformation functions, the metal thin films are assumed to have zero thickness ($t=0$), and two magnetic walls are placed on the dielectric--vacuum interfaces, i.e. the gaps of the CPW, assuming that the magnetic field is perpendicular to the vacuum-dielectric interfaces~\cite{Ghione1987}. As shown in Fig.~\ref{fig:3D2DRes_topMetal}(c), the resonator's cross-section can then be separated into two independent halves (top and bottom) by the metal surface of the control tier, from which different conformal mapping techniques are applied to the two halves. This results in parallel-plate waveguides with different widths and separations, see Appendix~\ref{appendix:Conformal transformations}. 

The total $L_l^{g,\text{conf}}$ and $C_l^{\text{conf}}$ values are therefore calculated as the parallel combination of the two halves. We have~\cite{Ghione1987,Gevorgian1995}
\begin{align}
    &\begin{aligned}
        L_l^{g,\text{conf}} = \left(1/L_l^{g,t}+1/L_l^{g,b}\right)^{-1} = \frac{\mu_0}{2}\left[\frac{K(k_s)}{K(k'_s)}+\frac{K(k_{1})}{K(k'_{1})}\right]^{-1}, 
        \label{eqn:Lg_theory_topMetal}
    \end{aligned}\\
    &\begin{aligned}
    C_l^{\text{conf}} & = C_l^{t} + C_l^{b}  \\
        & = 2\varepsilon_0\frac{K(k_{s})}{K(k'_{s})}+2\varepsilon_0\left[\frac{K(k_1)}{K(k'_1)} + (\varepsilon_r-1)\frac{K(k_2)}{K(k'_2)}\right],
    \label{eqn:C_theory_topMetal}
    \end{aligned}
\end{align}
where $t$ and $b$ represent the top and bottom halves of the CPW cross-section, respectively. Here $\mu_0$ is the vacuum permeability, $\varepsilon_0$ is the vacuum permittivity, $\varepsilon_r$ is the relative permittivity of the substrate of both tiers, and $K(k)$ is the complete elliptic integral of the first kind with modules
\begin{align}
    k_1 &= \frac{w_r}{w_r+2s_r}, \\
    k_2 &= \sinh\left[\frac{\pi w_r}{4h_{b}}\right]/\sinh\left[\frac{\pi (w_r+2s_r)}{4h_{b}}\right], \\
    k_{s} &= \tanh\left[\frac{\pi w_r}{4h_s}\right]/\tanh\left[\frac{\pi (w_r+2s_r)}{4h_s}\right], \\
    k_i' &= \sqrt{1-k_i^2}\,\,(i = 1,2,s).
\end{align}
A detailed derivation of (\ref{eqn:Lg_theory_topMetal}--\ref{eqn:C_theory_topMetal}) is shown in Appendix~\ref{appendix:Conformal transformations}.

We then compare $L_l^{g,\text{conf}}$ and $C_l^{\text{conf}}$ obtained by conformal mapping techniques with 2D FEM simulation under different inter-chip spacings $h_s$. We use $h_{b}=h_t=280\,\upmu$m, and $\varepsilon_r = 11.45$ for our high-resistivity silicon substrate of both tiers at cryogenic temperature~\cite{Krupka2006}. We approximate our aluminum superconducting films ($t=150\,$nm) using finite-thickness perfect conductors during the 2D FEM simulation. We use $w_r=s_r=12\,\upmu$m for the resonator's CPW geometry.

\begin{figure}[ht]
    \centering
    \includegraphics[width=1\linewidth]{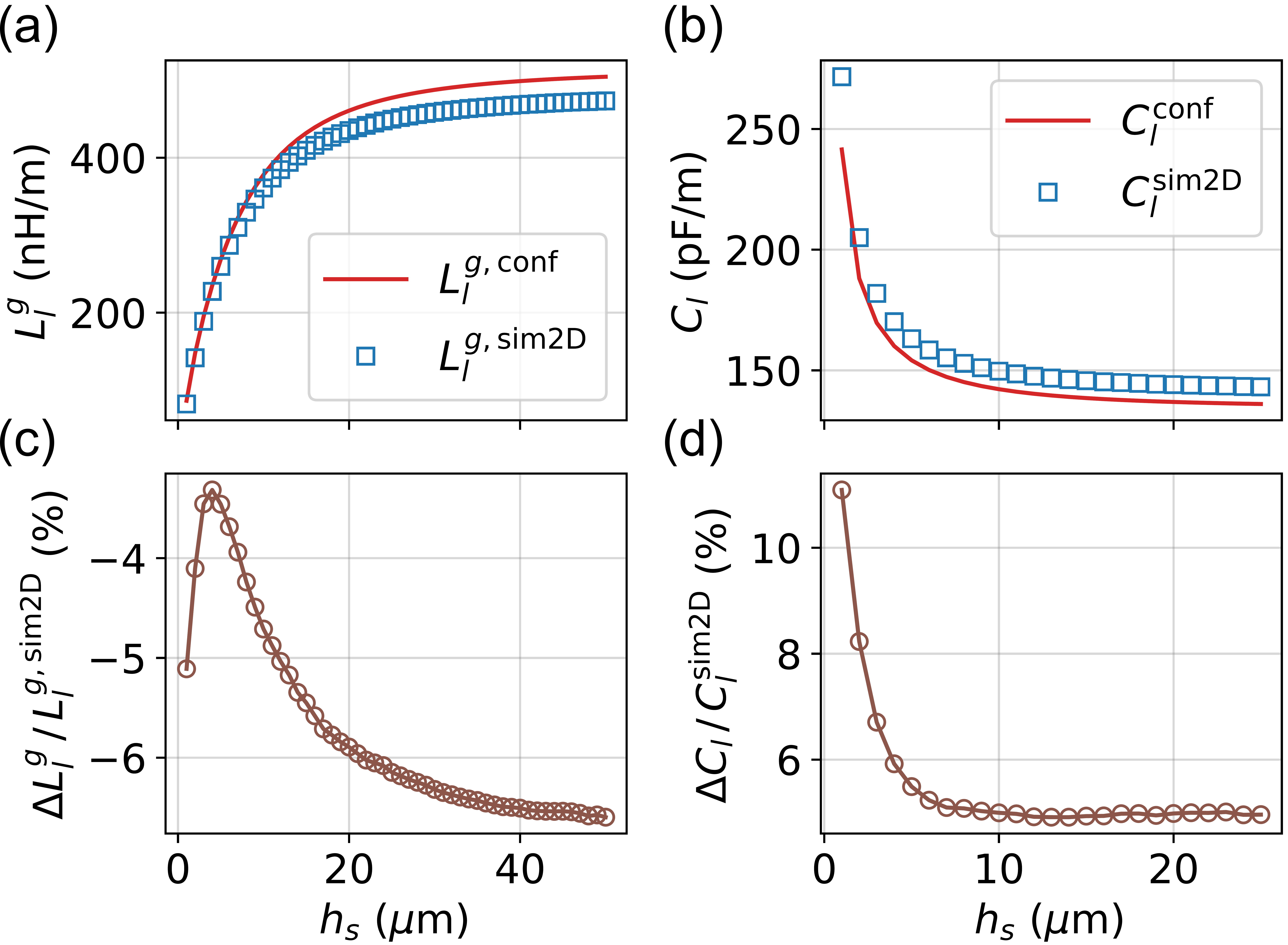}
    \caption{(a) Geometric inductance per unit length $L_l^g$ and (b) capacitance per unit length $C_l$ of the resonator's CPW cross-section obtained by conformal mapping calculation and 2D FEM simulation, respectively, under different inter-chip spacings $h_s$. (c--d) Their differences $\Delta L_l^g=L_l^{g,\text{sim2D}}-L_l^{g,\text{conf}}$ and $\Delta C_l=C_l^{\text{sim2D}}-C_l^{\text{conf}}$.}
    \label{fig:C_Lg_theory_2D_combined_topMetal}
\end{figure}

As shown in Fig.~\ref{fig:C_Lg_theory_2D_combined_topMetal}(a--b), when $h_s$ is small, $L_l^{g}$ and $C_l$ is heavily influenced by the top half part of the cross-section, governed by the factor $K(k_{s})/K(k'_{s})$ in~(\ref{eqn:Lg_theory_topMetal}) and~(\ref{eqn:C_theory_topMetal}). When $h_s$ increases (the opposing ground plane is moving away from the CPW resonator), its influence decreases and both conformal-mapping values and simulated values approach a limit as if the ground plane was absent ($h_s=\infty$). 

For comparison between the conformal mapping technique and 2D FEM simulation, we calculate $\Delta L_l^g=L_l^{g,\text{sim2D}}-L_l^{g,\text{conf}}$ and $\Delta C_l=C_l^{\text{sim2D}}-C_l^{\text{conf}}$ for each $h_s$, as shown in Fig.~\ref{fig:C_Lg_theory_2D_combined_topMetal}(c--d). We find that $\Delta L_l^g/L_l^{g,\text{sim2D}}$ and $\Delta C_l/C_l^{\text{sim2D}}$ stabilize at around -7\% and 5\% respectively for large $h_s$. This difference can be understood as the error of the zero-thickness assumption of the metal thin film, which is used during the conformal mapping transformation, compared with 2D FEM simulations which consider the edge effect of finite-thickness thin films\cite{kaiser2004electromagnetic}. When $h_s\leq4\,\upmu$m, the invalidity of the introduced magnetic wall on the resonator gap areas during conformal-mapping calculations is clearly shown, as $\Delta L_l^g/L_l^{g,\text{sim2D}}$ and $\Delta C_l/C_l^{\text{sim2D}}$ increase rapidly with smaller $h_s$. The cross-section of the CPW resonator in the flip-chip geometry therefore needs to be considered as a whole when $h_s$ is very small, rather than being separated into two halves and treated independently.

Having determined $L_l^g$ and $C_l$, we also need to know the kinetic inductance to calculate $f_r$. 
It is given by~\cite{Yoshida1992,Amini2022}
\begin{equation}
    L_l^k = \frac{\mu_0\lambda_m^2}{|I|^2}\cdot\int{J_z^2}\,dS,
    \label{eqn: Lk_int}
\end{equation}
where $J_z$ is the supercurrent density in the direction of the current flow, and the surface integral is over the cross-section of the thin film only. For a flip-chip geometry, the integral also includes the qubit tier's metal ground plane. Furthermore, $\lambda_m$ is the magnetic penetration depth of the superconductor, and $I$ is the total current injected into the CPW's center conductor.

To calculate (\ref{eqn: Lk_int}), we need to know $\lambda_m$ and $J_z$ over the whole thin film cross-section. We determine $J_z$ by electromagnetic simulation as in Ref.~\cite{Niepce2020}. In order to circumvent the complexity of evaluating $\lambda_m$ of a deposited aluminum thin film\cite{Miller1959,Tedrow1971,Cohen1968,Dressel2013,Mangin2016}, we directly compare measured resonant frequencies of six single-chip resonators with their simulated values and obtain a fitted $\lambda_m=83\,$nm, assuming the average frequency discrepancy between measurement and simulation is due to the omission of the kinetic inductance during the simulations. More details of the method to extract $\lambda_m$ is described in Appendix~\ref{appendix: lambda}. 

Fig.~\ref{fig:Lk_theory_2D_csp_topMetal} shows $L_l^k$ and its ratio to $L_l^{g,\text{sim2D}}$ under different $h_s$. Notice that $L_l^k$ has a minimum around $h_s=7\,\upmu$m. As the surface integral is on $J_z^2$, both the injected current on the center conductor and the return current on both tiers' ground planes contribute to $L_l^k$. The simulated distribution of $J_z$ shows that when $h_s$ is small, the return current concentrates on the qubit tier's metal directly facing the resonator's central conductor, and when $h_s$ is large, most of $J_z$ flows on the edges of the control tier's CPW structure\cite{Niepce2020, Amini2022}. When $h_s$ is at around $7\,\upmu$m, the distribution of $J_z$ is in-between the two extremes and has permeated the largest cross-section area of the thin film.

\begin{figure}[ht]
    \centering
    \includegraphics[width=1\linewidth]{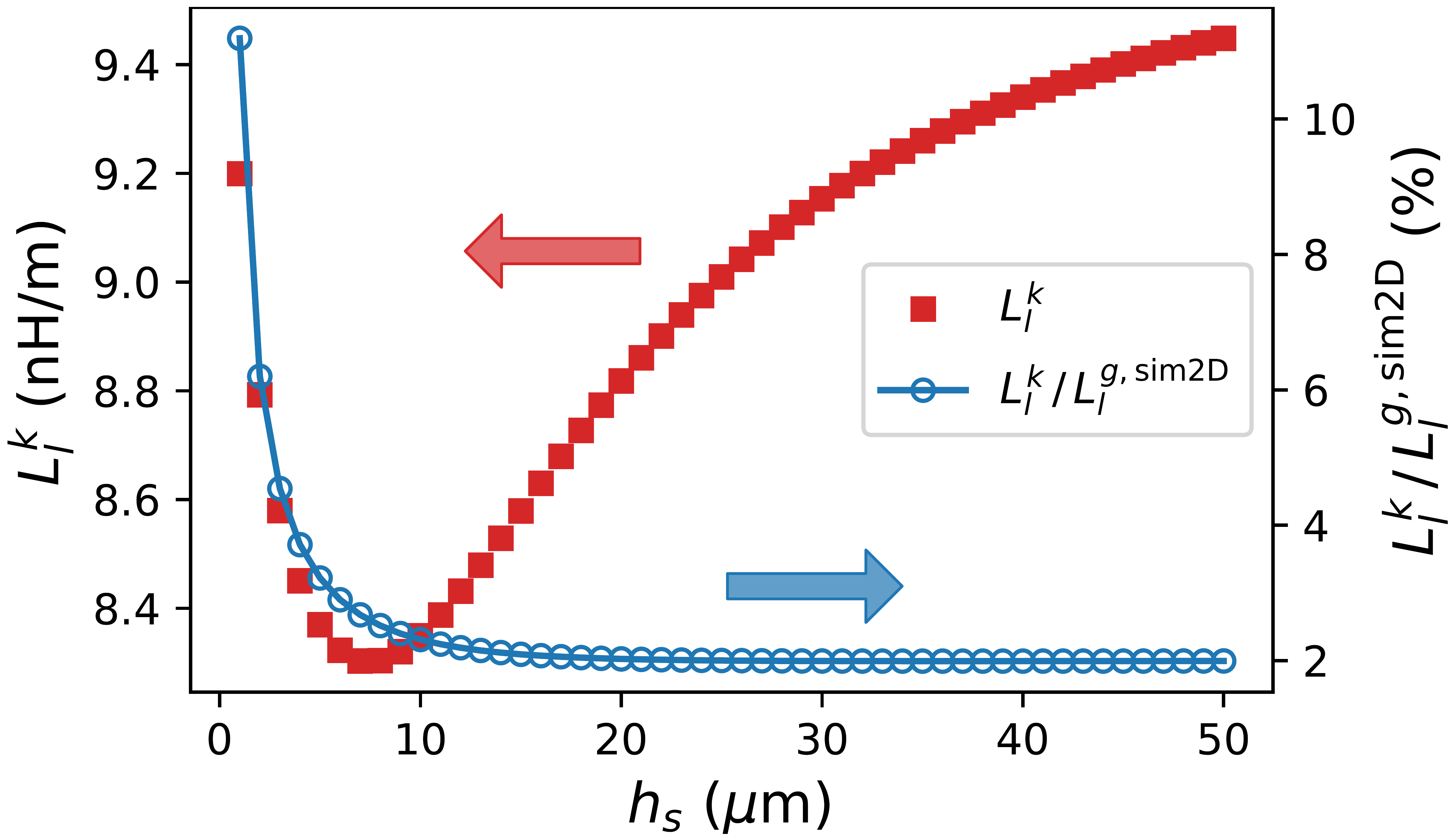}
    \caption{Simulated kinetic inductance per unit length $L_l^k$ and its ratio to the simulated geometric inductance per unit length $L_l^{g,\text{sim2D}}$ under different inter-chip spacings $h_s$. We use a magnetic penetration depth $\lambda_m=83\,$nm for the superconducting thin films during the simulation.}
    \label{fig:Lk_theory_2D_csp_topMetal}
\end{figure}

\subsection{Coupling quality factor}
The coupling quality factor $Q_c$ of the resonator quantifies its coupling rate to the feedline,  determining how fast the qubit state can be detected~\cite{Blais2021}. In addition, coupling to the feedline changes the resonator's characteristic impedance in the coupling area and causes a frequency shift $\delta f_r^c$~\cite{Besedin2018}. 

We use conformal mapping results~\cite{Besedin2018} to calculate the resonator's $Q_c$ and $\delta f_r^c$. Assuming the characteristic impedance of the feedline matches the input/output port ($50~\Omega$), we have
\begin{align}
    \frac{1}{Q_c} & = \frac{2\kappa^2\sin^2\theta}{\pi(2p-1)}, \label{eqn:Qc_theory_topMetal}
\end{align}
\begin{align}
    \delta f_r^c & = -\frac{c_l\sin\theta}{2\pi l_{tot}}\cdot \notag \\
    & \quad \left[\frac{\kappa^2(2\cos\psi+\cos\theta)}{2}+\frac{(Z_2-Z_r)\cos\psi}{Z_r}\right],\label{eqn:df_theory_topMetal}
\end{align}
with
\begin{align}
    \kappa &= - C_{rf}/\sqrt{C_{rr}C_{ff}}, \\
    c_l &= f_r\cdot4l_{tot}, \\
    Z_2 &= 1/\left(c_lC_{ff}\sqrt{1-\kappa^2}\right), \\
    \theta &= 2\pi l_c/(4l_{tot}), \\
    \psi &= 2\pi(l_c+2l_o^{e})/(4l_{tot}), 
\end{align}
where $f_r$ is the bare resonator frequency without feedline, and we take $p=1$ for its fundamental resonance. As shown in Fig.~\ref{fig:3D2DRes_topMetal}(b), $l_c$ is the length of the coupling part between resonator and feedline, including two additional 90-degree arcs at both ends of the coupling part to take into account the spurious coupling~\cite{Besedin2018}; $l_o^{e}$ is the effective length of the open part of the resonator, including the effective length of the coupling structure to the qubit; $Z_r$ is the characteristic impedance of the resonator; and $c_l$ is the speed of light within the resonator's CPW cross-section. The coupling capacitance ratio $\kappa$ and the impedance $Z_2$ of the resonator's coupling part are calculated from the capacitance between the resonator's and the feedline's center conductors $C_{rf}$, and their self capacitances ($C_{rr}$, $C_{ff}$). 

In order to obtain precise values of $C_{rf}$, $C_{rr}$, and $C_{ff}$ under different $h_s$, we conduct 2D FEM simulations at the cross-section of the resonator's coupling part to the feedline. Fig.~\ref{fig:3D2DRes_topMetal}(d) shows the cross-section of the resonator's coupling area where its center conductor is in parallel with the feedline.

\section{Comparison with 3D FEM simulation}
\label{section: Comparison with 3D FEM simulation}
To validate the accuracy of the resonator's $f_r$ and $Q_c$ obtained from 2D cross-sections, both parameters are compared to 3D FEM simulation~\cite{ansys3D} of a resonator as shown in Fig.~\ref{fig:3D2DRes_topMetal}(a) with $(l_s,l_c,l_o)=(3780.3, \, 425.7, \, 850.4)\,\upmu$m, where $l_s$ is the short part and $l_o$ is the open part of the resonator up to the center of the circular structure for qubit-resonator coupling. The circular structure has inner radius $R=29.4\,\upmu$m. Here $L_l^k$ is not included during the $f_r$ calculation in this comparison, because a perfect-\textit{E} boundary condition on the 2D zero-thickness sheets is used during the 3D FEM simulations to approximate the superconducting thin films\cite{Kosen2021}. We use $w_f=9\,\upmu$m and $s_f=10\,\upmu$m for the feedline's CPW geometry with the coupling gap $d=6\,\upmu$m during the $Q_c$ calculation.

To obtain a precise value of $l_{tot}$ in (\ref{eqn: fr_2D}), in addition to the resonator's designed length $l_r=l_s+l_c+l_o$, we also need to estimate the effective length of the circular structure at the open end of the resonator. We introduce two empirical coefficients ($\alpha_1$, $\alpha_2$) so that $l_{tot} = l_r + \alpha_1 R^2 + \alpha_2 R$. These coefficients are obtained by sweeping $R$ at a fixed $h_s$ and fitted with $f_r$ from 3D simulations. We obtain $\alpha_1 = 0.032\,\upmu$m$^{-1}$ and $\alpha_2 = 2.9$. We disregard the small effective length of the short end of the resonator~\cite{Simons2001}. More details are found in Appendix~\ref{appendix: eff length}. 

Fig.~\ref{fig:ResFreqQc_Cal_2D_vs_3D_topMetal}(a--b) shows the comparison between $f_r$ obtained by 3D FEM simulations ($f_r^{\text{sim3D}}$), 2D FEM simulations ($f_r^{\text{sim2D}}$), and conformal mapping techniques ($f_r^{\text{conf}}$) under different $h_s$ values. The cross-section calculations, $f_r^{\text{sim2D}}$ and $f_r^{\text{conf}}$, were calculated from their $L_l^g$ and $C_l$ values, with $\delta f_r^c$ added. In particular, $f_r^{\text{sim2D}}$ and $f_r^{\text{sim3D}}$ differ by 2\% over the whole $h_s$ range, and $f_r^{\text{conf}}$ is similarly accurate when $h_s\geq3\,\upmu$m.

Considering a typical readout resonator with $f_r=6\,$GHz, we therefore expect less than $100\,$MHz error between frequencies obtained by 2D cross-section and 3D FEM simulation. This error is acceptable in typical multiplexing readout situations~\cite{Heinsoo2018,Chen2022}. The benefit of the 2D cross-section method is that it substantially reduces the simulation time and required computer memory when designing resonators over a parameter split, or in large circuits. We show the comparison of required computational resources between 3D FEM simulation and 2D FEM simulation in Appendix~\ref{appendix: computing resource}---in our case a factor 1000. Furthermore, conformal-mapping techniques require only analytical calculations, reducing the need for extensive computational resources.

Fig.~\ref{fig:ResFreqQc_Cal_2D_vs_3D_topMetal}(c--d) also shows a difference of less than 20\% between $Q_c^{\text{cal2D}}$, obtained from a 2D cross-section, and that from 3D FEM simulation, $Q_c^{\text{sim3D}}$, when $h_s\geq7\,\upmu$m. When $h_s$ is smaller than 7$\,\upmu$m, the feedline may become less well matched to the input/output port.

\begin{figure}[ht]
    \centering
    \includegraphics[width=\linewidth]{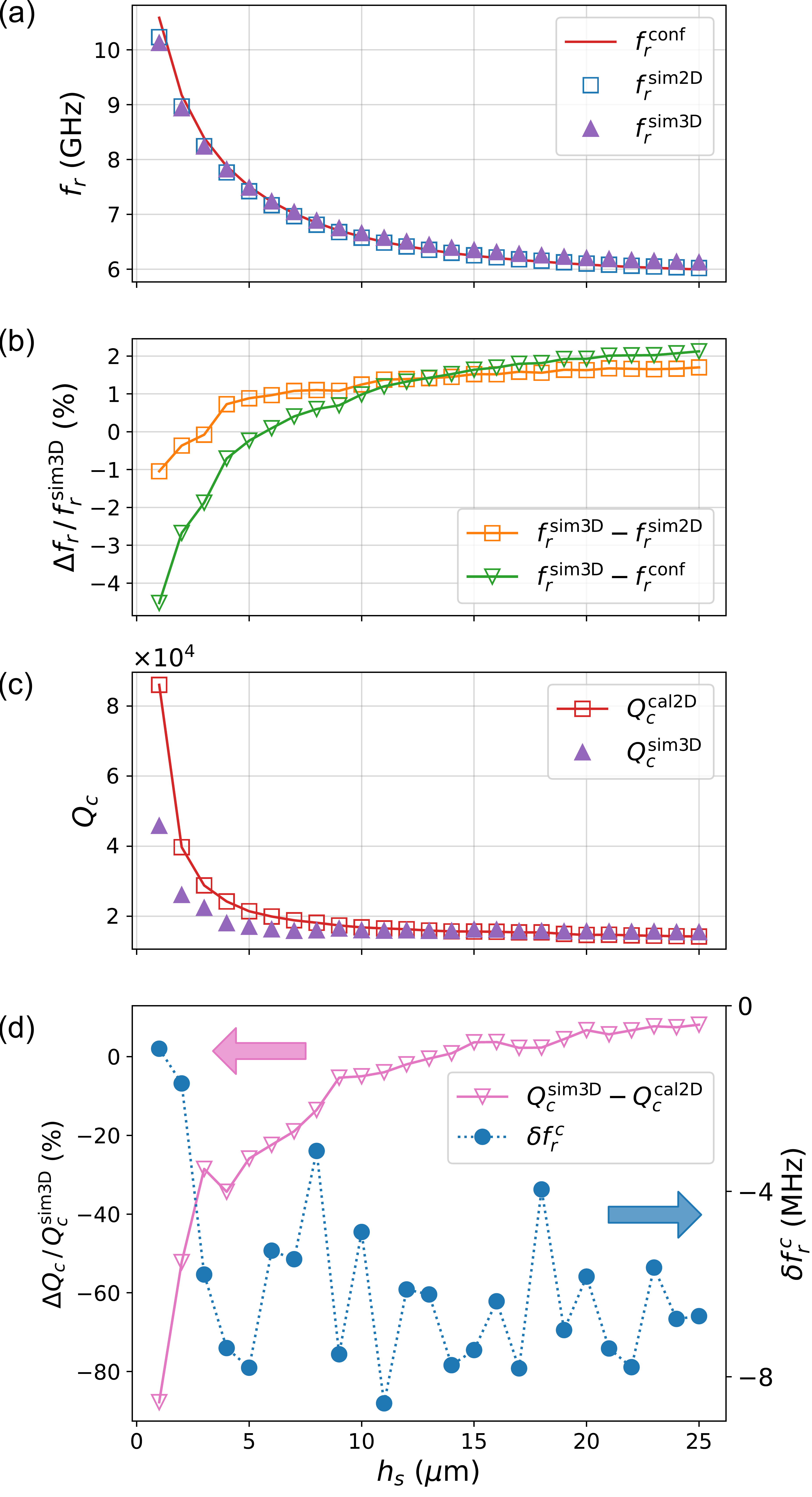}
    \caption{(a) Resonator frequencies obtained from conformal mapping calculation $f_r^{\text{conf}}$, 2D FEM simulation $f_r^{\text{sim2D}}$, and 3D FEM simulation $f_r^{\text{sim3D}}$ under different inter-chip spacings $h_s$. (b) Frequency difference $\Delta f_r$ between 3D and 2D FEM simulation, and between 3D FEM simulation and conformal mapping calculation. (c) Resonator coupling quality factor obtained from 2D cross-sections $Q_c^{\text{cal2D}}$ and 3D FEM simulation $Q_c^{\text{sim3D}}$ and (d) their difference, together with the coupling-induced frequency shift $\delta f_r^c$ under different $h_s$. $\delta f_r^c$ is calculated using $f_r^{\text{sim2D}}$ for the bare resonator frequency. }
    \label{fig:ResFreqQc_Cal_2D_vs_3D_topMetal}
\end{figure}

\section{Comparison with measurement}
We now examine our 2D cross-section techniques by calculating the resonant frequencies and coupling quality factors of fifteen resonators by means of conformal transformation and 2D FEM simulation, $f_r^{\text{conf+KI}}$, $f_r^{\text{sim2D+KI}}$, and $Q_c^{\text{cal2D}}$ (including simulated kinetic inductance $L_l^k$), and comparing them with experimentally measured values. 
The resonators were fabricated and packaged within a multi-qubit flip-chip-integrated quantum processor as reported by Kosen and Li~\cite{Kosen2021}. They were measured at millikelvin temperature, and their parameters $f_r^{\text{meas}}$ and $Q_c^{\text{meas}}$ were determined from the forward transmission scattering parameter $S_{21}$ in the high-power regime, where these resonators are decoupled from their corresponding qubits~\cite{Boissonneault2010}. 

\begin{figure}[t]
    \centering
    \includegraphics[width=1\linewidth]{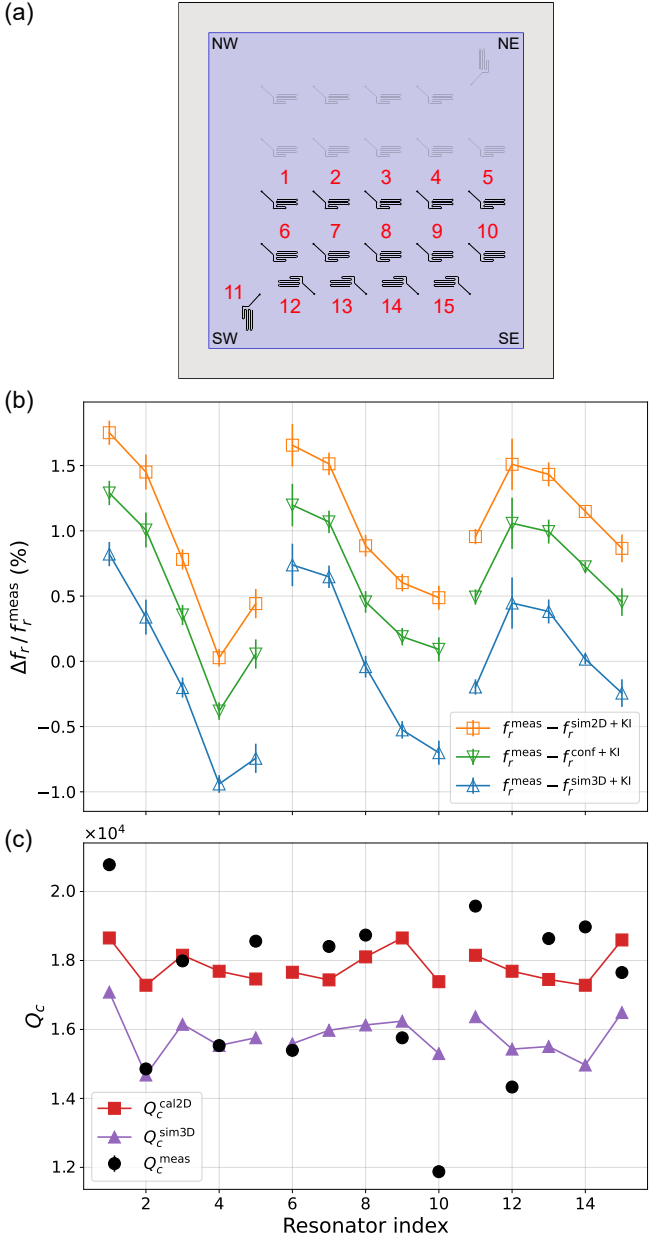}
    \caption{Comparison between 2D cross-section calculation and measurement results of 15 resonators within a flip-chip quantum processor. (a) Illustration of the positions of 15 resonators within the processor. Gap measurement by scanning electron microscopy at the processor's four corners yielded [NW, NE, SW, SE] = [8.3, 9.3, 8.3, 8.8]$\,\upmu$m. (b) Frequency difference between measurements $f_r^{\text{meas}}$, conformal mapping calculations $f_r^{\text{conf+KI}}$, 2D FEM simulations $f_r^{\text{sim2D+KI}}$ and 3D FEM simulations $f_r^{\text{sim3D+KI}}$, including kinetic inductance (KI). (c) Coupling quality factor obtained from measurements $Q_c^{\text{meas}}$, 2D cross-section calculations $Q_c^{\text{cal2D}}$ and 3D FEM simulations $Q_c^{\text{sim3D}}$.}
    \label{fig:ResFreq_Cal_2D_vs_Meas_topMetal}
\end{figure}

To obtain the value of $h_s$ for each of the 15 resonators within one flip-chip-integrated processor, we used the gap values, measured by scanning electron microscopy, at the chip's four corners~\cite{Kosen2021} and inferred the tilt of the inter-chip distance using bilinear interpolation (assuming flat surfaces). Fig.~\ref{fig:ResFreq_Cal_2D_vs_Meas_topMetal}(a) shows the locations of the 15 resonators.

Fig.~\ref{fig:ResFreq_Cal_2D_vs_Meas_topMetal}(b) shows a comparison between the calculated and measured $f_r$ of these 15 resonators. Here $f_r^{\text{conf+KI}}$ and $f_r^{\text{sim2D+KI}}$ deviate from the measured frequencies ($f_r^{\text{meas}}$) by less than 2\%, demonstrating the accuracy of our 2D cross-sectional calculation method. The inclusion of the kinetic inductance has shifted up these frequency differences by about 1.2\%, resulting in a deviation between $f_r^{\text{meas}}$ and $f_r^{\text{sim3D+KI}}$ to be around 0\%. We observe the monotonic change of the frequency differences at each row of resonators, indicating that there is a residual chip tilt that has not been compensated by the bilinear interpolation of the inter-chip distance from the chip's four corners.

Fig.~\ref{fig:ResFreq_Cal_2D_vs_Meas_topMetal}(c) shows that across 15 resonators, $Q_c^{\text{cal2D}}$ and $Q_c^{\text{sim3D}}$ are kept at certain values, as all resonators' coupling lengths $l_c$ and coupling gaps $d$ are designed to be the same. $Q_c^{\text{cal2D}}$ and $Q_c^{\text{sim3D}}$ differ by around 13\%, cf.\@ Fig.~\ref{fig:ResFreqQc_Cal_2D_vs_3D_topMetal}(c) for $h_s\sim8~\upmu$m. The measured coupling quality factors $Q_c^{\text{meas}}$ oscillate across 15 resonators, resulting in deviation from $-20\%$ to $+10\%$ to either $Q_c^{\text{cal2D}}$ or $Q_c^{\text{sim3D}}$. We attribute such oscillation to impedance mismatch between feedline and input/output port at wire-bond interconnections to the printed circuit board.

\section{Inter-chip-spacing insensitive flip-chip resonator design}
\label{seciton: inter-chip-spacing insensitive design of flip-chip resonator}

In Fig.~\ref{fig:ResFreqQc_Cal_2D_vs_3D_topMetal}(a) we see that the resonator frequency $f_r$ is sensitive to the change of inter-chip spacing $h_s$, especially when $h_s$ is below 10$\,\upmu$m, where 1$\,\upmu$m variation of $h_s$ can result in larger than 100$\,$MHz change of $f_r$. We note that the rapid increase of $f_r$ for decreasing $h_s$ is due to a rapid decrease of $L_l^g$. 
In this section, we propose a method to reduce the sensitivity of $f_r$ to $h_s$ by partly removing the metal ground plane facing the CPW resonator on the opposing chip. We first use conformal transformation to determine ${L_l^g}'$ and $C_l'$ of the resonator's cross-section in the case where no metal is facing the CPW line, i.e. the CPW line is directly facing the dielectric substrate of the qubit tier. We then calculate the optimal ratio between the resonator's dielectric-facing length to its total length, such that at a certain range of $h_s$, the sensitivity of $f_r$ to $h_s$ is minimized.

\subsection{Resonator CPW facing dielectric}
\label{seciton: Resonator exposed to substrate of top chip}

We use conformal mapping techniques to obtain ${L_l^{g,\text{conf}}}'$ and ${C_l^{\text{conf}}}'$ of the cross-section when the CPW resonator faces the dielectric substrate of the opposing chip. The resulting equations are 

\begin{align}
    &\begin{aligned}
        {L_l^{g,\text{conf}}}' = \frac{\mu_0}{4}\frac{K(k'_1)}{K(k_1)},
        \label{eqn:Lg_2D_topSi}
    \end{aligned}\\
    &\begin{aligned}
        {C_l^{\text{conf}}}' & = \frac{2\varepsilon_0}{\left(\varepsilon_r\frac{K(k_1)}{K(k'_1)}\right)^{-1}+ \left(\frac{\varepsilon_r}{\varepsilon_r-1}\frac{K(k_{s})}{K(k'_{s})}\right)^{-1}} \\
        & +2\varepsilon_0\left[\frac{K(k_1)}{K(k'_1)}+(\varepsilon_r-1)\frac{K(k_2)}{K(k'_2)}\right].
       \label{eqn:C_2D_topSi}
    \end{aligned}
\end{align}
To derive the capacitance per unit length, as in Appendix~\ref{appendix:Conformal transformations}, we separate the capacitance contribution of the vacuum from that of the qubit tier's substrate at the top half of the cross-section. However, since the CPW metal is closer to the inter-chip vacuum than the qubit tier's substrate, their capacitance contributions are taken in series rather than in parallel~\cite{Ghione2003}.

\begin{figure}[ht]
    \centering
    \includegraphics[width=\linewidth]{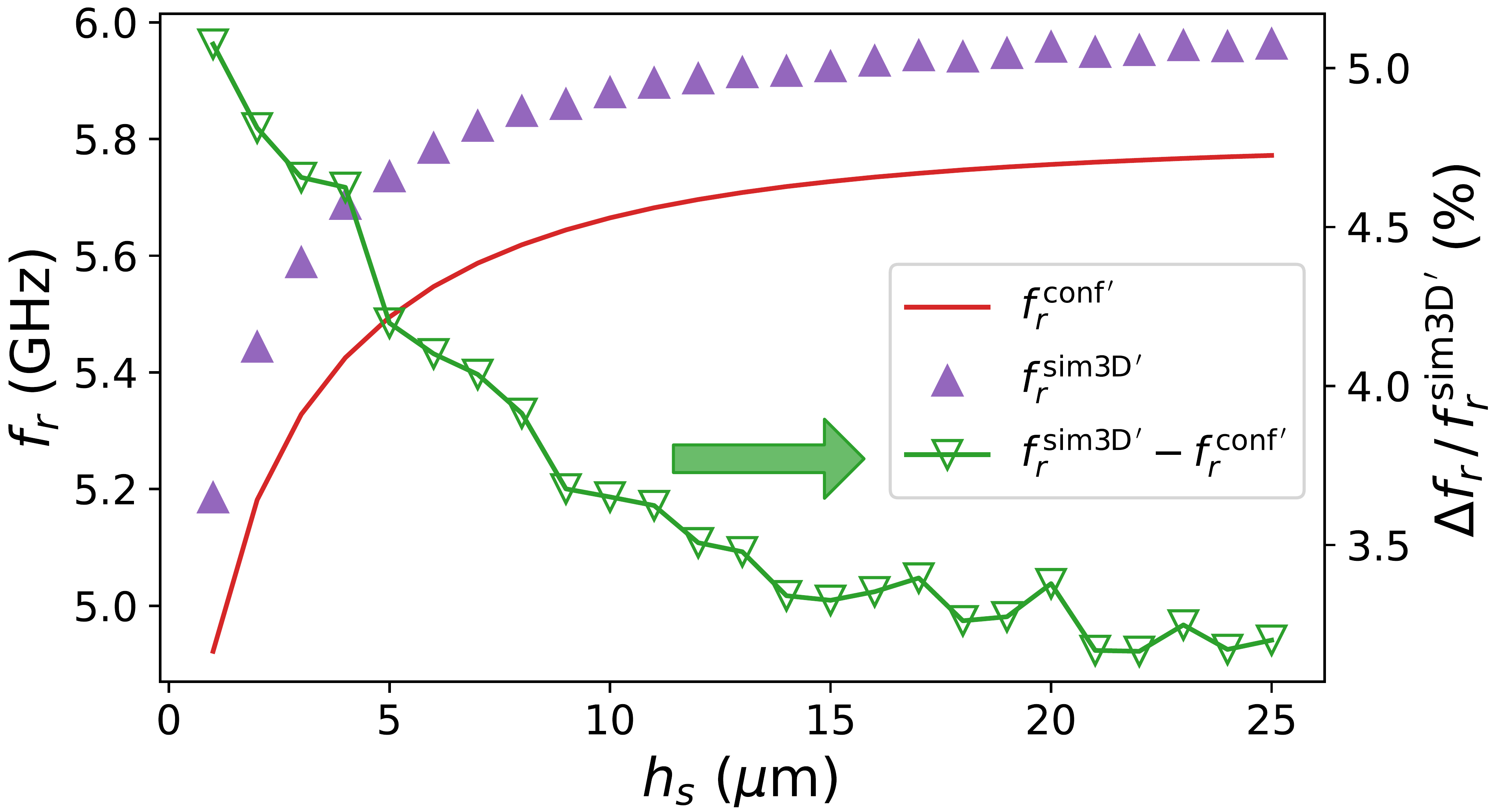}
    \caption{Resonator frequencies obtained from conformal mapping calculation ${f_r^{\text{conf}}}'$ and 3D FEM simulation ${f_r^{\text{sim3D}}}'$ under different inter-chip spacings $h_s$ when the resonator's CPW line is facing a dielectric substrate on the opposing chip, together with the frequency difference between 3D FEM simulation and conformal mapping calculation. }
    \label{fig:ResFreq_3D_topSi}
\end{figure}

Fig.~\ref{fig:ResFreq_3D_topSi} shows ${f_r^\text{{conf}}}'$ and ${f_r^\text{{sim3D}}}'$ when the resonator's CPW line faces the dielectric. ${f_r^\text{{sim3D}}}'$ is obtained using the same resonator as in Section \ref{section: Comparison with 3D FEM simulation}, but with the entire metal film removed from the opposing chip. ${f_r^\text{{conf}}}'$ is calculated using ${L_l^{g,\text{conf}}}'$ and ${C_l^{\text{conf}}}'$, with an added coupling-induced frequency shift: it gives quantitatively the same result as ${f_r^\text{{sim3D}}}'$, differing by 3\% to 5.5\% under different $h_s$.

Using the same participation ratio simulation setting as in~\cite{Kosen2021}, we also simulate the participation ratio of lossy dielectric interfaces and compare the resulting Q-factors of the resonator's cross-section when the CPW line faces either a metal or a dielectric. See Appendix~\ref{appendix: participation ratio simulation} for details.

\subsection{Ground-plane cutout to reduce the frequency dependence on inter-chip-spacing}
Since in these two scenarios (CPW facing either metal or dielectric, cf.\@ Fig.~\ref{fig:ResFreqQc_Cal_2D_vs_3D_topMetal}(a) and Fig.~\ref{fig:ResFreq_3D_topSi}), the resonator's $f_r$ exhibits the opposite response to a change of inter-chip spacing $h_s$, we can use this to render $f_r$ insensitive to small $h_s$ variations around a chosen value within our typical range $6\,\upmu\mathrm{m}<h_s<10\,\upmu$m~\cite{Kosen2021}. 

We define $\gamma$ as the ratio between the resonator length facing dielectric substrate and the resonator's total length $l_{tot}$. We can approximate the effective inductance and capacitance per unit length as
\begin{align}
    L_l^{e} &= (1-\gamma) (L^{g,\text{conf}}_l+L_l^k) + \gamma ({L_l^{g,\text{conf}}}'+{L_l^k}'), \\
    C_l^{e} &= (1-\gamma) C_{l}^{\text{conf}} + \gamma {C_l^{\text{conf}}}',
\end{align}
where ${L_l^k}'$ is the kinetic inductance per unit length when the CPW faces dielectric.

To find the optimal value $\gamma_{\text{opt}}$ of minimal sensitivity to $h_s$ variation, we use a cost function $F(\gamma)$ independent of the total length:
\begin{equation}
    F(\gamma) = \sum_{i=1}^N\left|\frac{\partial(L_l^{e}C_l^{e})^{-1/2}}{\partial h_s^i}\right|,
\end{equation}
where $h_s^i=h_s^1+(i-1)(h_s^N-h_s^1)/N$ with $N\in\mathbb{N}^*$, and $\gamma_{\text{opt}}$ is the value that minimizes $F(\gamma_{\text{opt}})$ in the range $h_s\in[h_s^1,h_s^N]$.

\begin{figure}[ht]
    \centering
    \includegraphics[width=\linewidth]{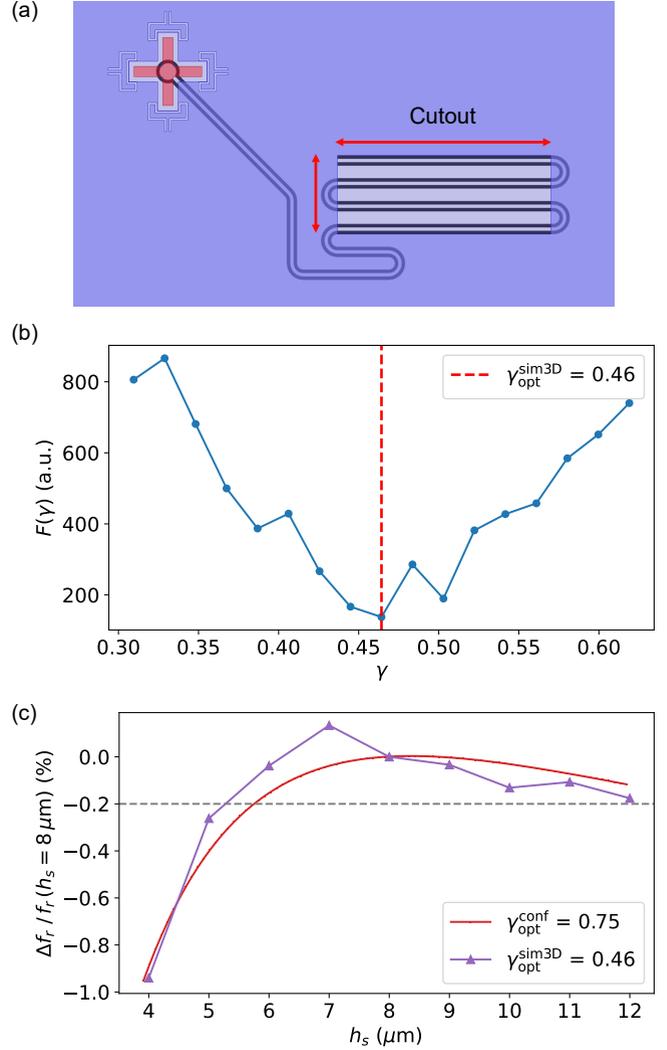}
    \caption{(a) Illustration of a $\lambda/4$ resonator facing a partial cut-out of the metal ground plane on the opposing chip. (b) Cost function $F(\gamma)$ calculated from 3D FEM simulations of the resonator with $(h_s^N-h_s^1)/N=1\,\upmu$m. The optimal coverage ratio $\gamma_{\text{opt}}^{\text{sim3D}}=0.46$ is found at the minimum of $F(\gamma)$. (c) Resonator frequency variation around $h_s=8\,\upmu$m for $\gamma = \gamma_{\text{opt}}$, obtained from conformal mapping calculation and 3D FEM simulation, respectively.}
    \label{fig:FC7_Res3D_combineMetalSi}
\end{figure}

We obtain $\gamma_{\text{opt}}^{\text{conf}}=0.75$ for $h_s\in[6,10]\,\upmu$m for large enough $N$. We ignore the kinetic inductance during the calculation, using the fact that at large $h_s$, $L^{g,\text{conf}}_l={L_l^{g,\text{conf}}}'$ and $L^{k}_l={L_l^k}'$, and when $h_s\geq6\,\upmu$m, $L_l^k/L^{g,\text{conf}}_l\approx{L_l^k}'/{L_l^{g,\text{conf}}}'$ (Fig.~\ref{fig:Lk_theory_2D_csp_topMetal}). As shown in Fig.~\ref{fig:FC7_Res3D_combineMetalSi}(c), the deviation of $f_r$ from its value at $h_s=8\,\upmu\text{m}$, within this range, can be less than 0.2\%.

We employ this principle to the actual resonator design as in Fig.~\ref{fig:3D2DRes_topMetal} to validate our calculation. For simplicity, we cut out a rectangular area on the opposing chip's metal ground plane, so that part of the resonator's CPW line faces the substrate, as shown in Fig.~\ref{fig:FC7_Res3D_combineMetalSi}(a). We found that $\gamma_{\text{opt}}^{\text{sim3D}}=0.46$ for the actual resonator and obtained a similar deviation of $f_r$ around $h_s=8\,\upmu\text{m}$ as the calculation (Fig.~\ref{fig:FC7_Res3D_combineMetalSi}(b-c)). We attribute the discrepancy between $\gamma_{\text{opt}}^{\text{conf}}$ and $\gamma_{\text{opt}}^{\text{sim3D}}$ to the meandering structure of the resonator, supported by the additional simulation result that $\gamma_{\text{opt}}^{\text{sim3D}}=\gamma_{\text{opt}}^{\text{conf}}$ if the resonator is realized by a straight CPW line instead.

\section{Conclusion}
In conclusion, we have shown that analytic and numerical simulation of a 2D cross-section of a coplanar waveguide superconducting resonator in flip-chip geometry can predict the resonant frequency within 2\% of 3D-simulated and measured values, after considering the effective length correction due to the resonator ends. We also determined the coupling quality factor within 20\% by 2D simulation. 

The 2D cross-sectional method is considerably more computationally efficient than conventional 3D FEM simulations. The 3 orders of magnitude speed-up will become particularly useful for large-scale quantum processor design containing a large number of qubits and resonators. The conformal mapping analytic equations and 2D simulation steps can be easily integrated into the process design kit workflow of superconducting quantum processor design, e.g., using newly developed tools like Qiskit metal. This method can also be generalized to other components implemented using CPW transmission lines, e.g., couplers and Purcell filters.

In addition, we have proposed a resonator design that is insensitive to inter-chip-spacing variations. 

\appendices

\section{Conformal transformations for flip-chip CPW cross-section}
\label{appendix:Conformal transformations}
In this appendix we will use conformal mapping techniques to derive the geometric inductance and capacitance per unit length of the CPW cross-section in flip-chip geometry when the CPW line faces a metal ground plane on the opposing chip.

To simplify our conformal transformation functions such that they are feasible to calculate analytically, we assume zero thickness of the metal thin films on both tiers. We also assume that the chip spacing $h_s$ is large enough, such that the magnetic field around the CPW center conductor is perpendicular to the vacuum--dielectric interfaces at gap areas of the CPW. We are then able to cover the vacuum--dielectric interfaces with magnetic walls and separate the CPW cross-section into two halves. See Fig.~\ref{fig:3D2DRes_topMetal}(c).

We then apply conformal transformations to the separated two halves independently, resulting in different parallel-plate waveguides
with different widths $W$ and separations $H$. The geometric inductance and capacitance per unit length of each parallel-plate waveguide are therefore calculated using
\begin{align}
    L_l^{g} &= \mu_0\frac{H}{W},\\
    C_l &= \varepsilon^p_{r}\varepsilon_0\frac{W}{H},
\end{align}
where $\varepsilon^p_{r}$ is the relative permittivity of the dielectric medium between two plates after the conformal transformation. We use different conformal mapping techniques for each half. The contributions of two halves to the total geometric inductance and capacitance per unit length of the CPW cross-section are calculated as a parallel combination. 

\subsection{The bottom half}
Now we will show how to do the conformal transformation to each half of the cross-section. We start with the bottom half in Fig.~\ref{fig:3D2DRes_topMetal}(c). This half can be seen as a conventional CPW cross-section having a dielectric substrate with relative permittivity $\varepsilon_r$. 

We first calculate the geometric inductance per unit length of the bottom-half cross-section. We replace the substrate with the vacuum since their permeability are the same. We first put the cross-section into a complex $z$-plane. Because all the metals are in the horizontal direction, we can put the metals along with the Re[$z$]-axis and directly apply the Christoffel-Schwartz transformation~\cite{schinzinger2003} to conformally map these metals into a parallel-plate waveguide~\cite{Wen1969} in the complex $w$-plane. The transformation function we use is
\begin{equation}
    w(z) = A_1\int^z_0\frac{dz}{\sqrt{(z-z_B)(z-z_C)(z-z_D)(z-z_E)}}+A_2.
    \label{eqn: conformal_CPW_func}
\end{equation}
Here $A_1$ and $A_2$ are constants that determine the scaling and translation of the transformed geometry, and $z_i\,(i = B,C,D,E)$ are the positions of the end points of the metals on the Re[$z$]-axis. Choosing the center of the CPW's center conductor as the zero position of the Re[$z$]-axis, we have $-z_B=z_E=(w_r+2s_r)/2$ and $-z_C=z_D=w_r/2$.

The geometry in the $z$-plane is now conformally mapped to the $w$-plane. As a result, the CPW center conductor and the ground plane (two infinite points viewed as connected) in the $z$-plane are transformed into the two plates with equal width and separated in parallel, forming a parallel-plate waveguide with vacuum in-between the plates.

The width and the height of this parallel-plate waveguide are calculated as
\begin{align}
   W_b^{\text{vac}} &= |w_D-w_C| = |A_1|2K(k_1), \\
   H_b^{\text{vac}} &= |w_E-w_D| = |A_1|K(k_1'),
\end{align}
where $K(k)$ is the complete elliptic integral of the first kind with modules $k_1 = z_D/z_E = w_r/(w_r+2s_r)$ and $k_1'=\sqrt{(1-k_1^2)}$.

The geometric inductance per unit length of this parallel-plate waveguide is obtained as
\begin{equation}
    L_l^{g,b} = \mu_0\frac{H_b^{\text{vac}}}{W_b^{\text{vac}}} = \frac{\mu_0}{2}\frac{K(k_1')}{K(k_1)}.
\end{equation}

To calculate the capacitance per unit length of the bottom-half cross-section, we can further view this cross-section as two cross-sections in a parallel combination, in which the first has the vacuum below the metals, whereas the second has a finite-thickness dielectric substrate with relative permittivity $\varepsilon_r - 1$. We can then calculate the capacitance contributions from the vacuum and the substrate separately. We refer to~\cite{Carlsson1999} for an illustration of this separation.

The capacitance per unit length of the cross-section with vacuum below the metals can be calculated using the same conformal transformation function (\ref{eqn: conformal_CPW_func}) as in the calculation of geometric inductance per unit length. Therefore we have
\begin{equation}
    C_l^{b,\text{vac}} = \varepsilon_0\frac{W_b^{\text{vac}}}{H_b^{\text{vac}}} = 2\varepsilon_0\frac{K(k_1)}{K(k_1')}.
\end{equation}

\begin{figure}[ht]
    \centering
    \includegraphics[width=\linewidth]{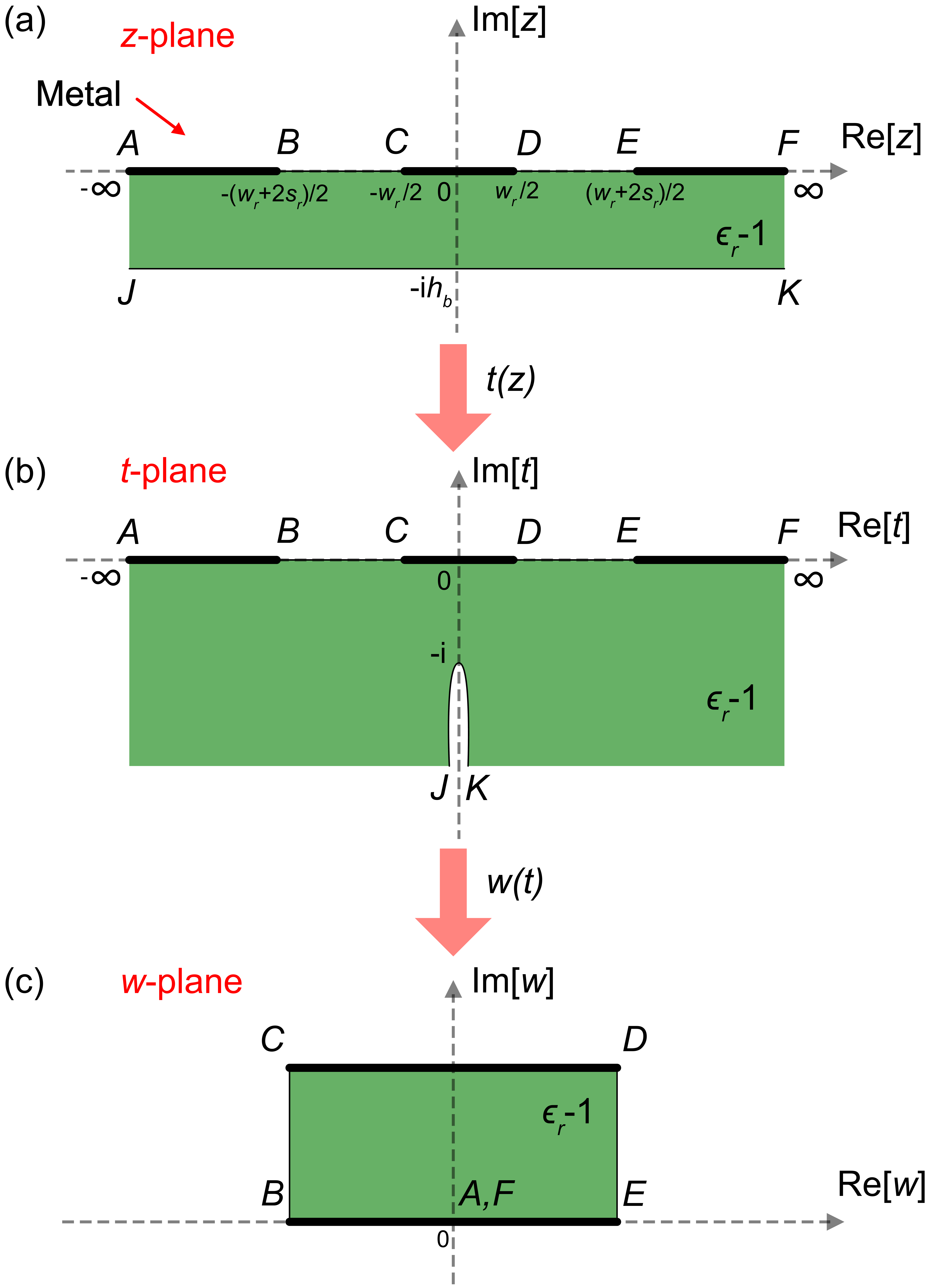}
    \caption{Conformal transformations for the bottom half of the flip-chip CPW cross-section when the substrate has relative permittivity $\varepsilon_r-1$. (a) Original geometry. (b) Intermediate geometry in the $t$-plane. (c) Mapped to a parallel-plate waveguide.}
    \label{fig:conformal_bottom_half}
\end{figure}

For the cross-section having a finite-thickness substrate with relative permittivity $\varepsilon_r - 1$, we can first do an intermediate transformation such that the substrate becomes infinitely thick to resemble the vacuum case above. We map the cross-section from the $z$-plane to the $t$-plane with function 
\begin{equation}
    t(z) = \sinh{\left[\frac{\pi z}{2h_b}\right]}.
\end{equation}
From the $t$-plane, we repeat the same Christoffel-Schwartz transformation but replace the variable notations in (\ref{eqn: conformal_CPW_func}) from $z$ to $t$. Fig.~\ref{fig:conformal_bottom_half} shows the two consecutive conformal transformations.

Thus we obtain 
\begin{equation}
    C_l^{b,\text{sub}} = (\varepsilon_r-1)\varepsilon_0\frac{W_b^{\text{sub}}}{H_b^{\text{sub}}} = 2(\varepsilon_r-1)\varepsilon_0\frac{K(k_2)}{K(k_2')},
\end{equation}
with modules
\begin{align}
    &\begin{aligned}
        k_2 = t_D/t_E & = \sinh{\left[\frac{\pi z_D}{2h_b}\right]}/\sinh{\left[\frac{\pi z_E}{2h_b}\right]} \\
           & = \sinh\left[\frac{\pi w_r}{4h_{b}}\right]/\sinh\left[\frac{\pi (w_r+2s_r)}{4h_{b}}\right],
    \end{aligned} \\
    & k_2' = \sqrt{1-k_2^2}.
\end{align}

Therefore, the geometric inductance and capacitance per unit length of the bottom half of the CPW cross-section are
\begin{align}
    & L_l^{g,b} = \frac{\mu_0}{2}\frac{K(k_1')}{K(k_1)}, \\
    & \begin{aligned}
        C_l^{b} &= C_l^{b,\text{air}} + C_l^{b,\text{sub}} \\
        &= 2\varepsilon_0\left[\frac{K(k_1)}{K(k_1')} + (\varepsilon_r-1)\frac{K(k_2)}{K(k_2')}\right].
    \end{aligned}
\end{align}

\subsection{The top half}
The top half of the CPW cross-section will be treated differently compared to the bottom half, as the top-half cross-section has a metal ground plane from the qubit tier on top of the CPW line. Since the qubit tier's metal ground plane blocks all the electromagnetic field generated from the CPW line, we can view the qubit tier's substrate as absent and replace it with vacuum. 

To simplify the calculation, we exploit the symmetry of the top-half cross-section and only calculate the geometric inductance and capacitance per unit length of the cross-section at the $z$-plane's real positive part (zero position is at the center of CPW's center conductor). The resultant two identical parallel-plate waveguides mapped from the real positive and negative parts are also treated as a parallel combination. 

\begin{figure}[ht]
    \centering
    \includegraphics[width=\linewidth]{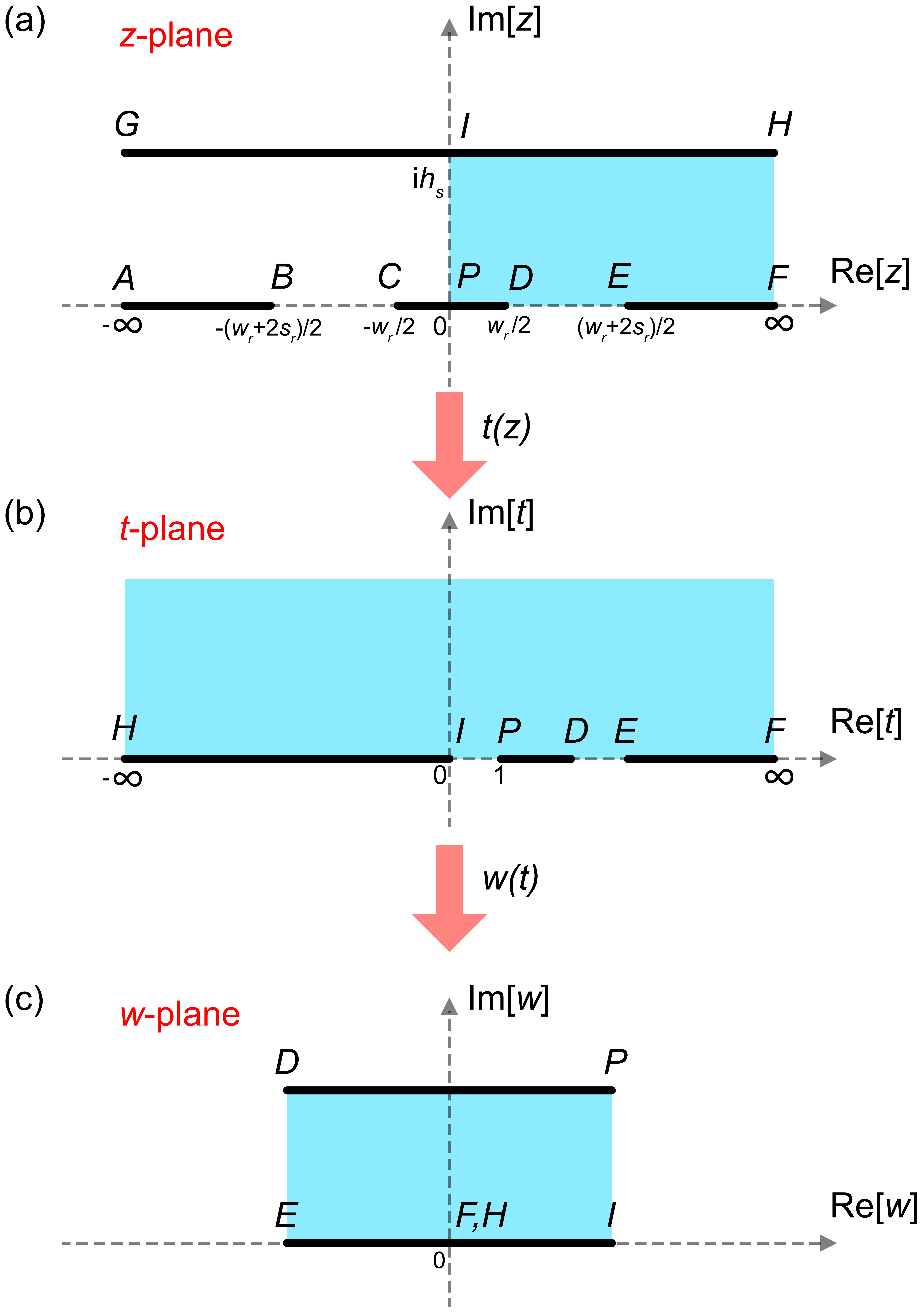}
    \caption{Conformal transformations for the top half of the flip-chip CPW cross-section. The transformed region is painted in cyan. (a) Original geometry. (b) Intermediate geometry in the $t$-plane. (c) Mapped to a parallel-plate waveguide.}
    \label{fig:conformal_top_half}
\end{figure}

We do two consecutive conformal transformations to map the real positive part of the top-half cross-section into a parallel-plate waveguide. We first map the qubit tier's half of the metal ground plane in the $z$-plane to the Re[$t$]-axis, and then use the Christoffel-Schwartz transformation to map the geometry to a parallel-plate waveguide in the $w$-plane~\cite{Gevorgian1995}. Fig.~\ref{fig:conformal_top_half} shows the two conformal transformations.

The first transformation function we use is
\begin{equation}
    t(z) = \cosh^2{\left[\frac{\pi z}{2h_b}\right]}.
\end{equation}

In the $z$-plane, we define $z_P=0$, $z_I=ih_s$, $z_D=w_r/2$ and $z_E=(w_r+2s_r)/2$. In the $t$-plane these points are mapped into $t_P=1$, $t_I=0$, $t_D=\cosh^2{\left[\pi w_r/4h_b\right]}$ and $t_E=\cosh^2{\left[\pi (w_r+2s_r)/4h_b\right]}$.

The second transformation function we use is
\begin{equation}
    w(t) = A_1 F(\varphi,k_s)+A_2,
\end{equation}
where $F(\varphi,k_s)$ is the elliptic integral of the first kind with
\begin{align}
    F(\varphi,k_s) &= \int^{\sin{\varphi}}_0 \frac{d\tau}{\sqrt{(1-k_s^2\tau^2)(1-\tau^2)}}, \\
    \sin{\varphi} &= \sqrt{\frac{(t-t_E)t_D}{(t-t_D)t_E}}, \\
    &\begin{aligned}
       k_s &= \sqrt{\frac{t_E(t_D-t_P)}{t_D(t_E-t_P)}} \\
        &= \tanh\left[\frac{\pi w_r}{4h_s}\right]/\tanh\left[\frac{\pi (w_r+2s_r)}{4h_s}\right].
    \end{aligned}
\end{align}

After the second transformation, the real positive part of the top-half cross-section is now mapped to a parallel-plate waveguide, with the width and the height
\begin{align}
   W_t &= |w_I-w_E| = |A_1|K(k_s), \\
   H_t &= |w_P-w_I| = |A_1|K(k_s'),
\end{align}
where we have used the relations
\begin{align}
    & F(\frac{\pi}{2},k_s) = K(k_s), \\
    & F(\arcsin{\frac{1}{k_s}},k_s) = K(k_s) + iK(k_s'), \\
    & k_s' = \sqrt{(1-k_s^2)}.
\end{align}

After combining the same results from the real negative part of the top-half cross-section, the geometric inductance and capacitance per unit length of the top half of the CPW cross-section are
\begin{align}
    & L_l^{g,t} = \frac{\mu_0}{2}\frac{H_t}{W_t} = \frac{\mu_0}{2}\frac{K(k_s')}{K(k_s)}, \\
    & C_l^{t} = 2\varepsilon_0\frac{W_t}{H_t} = 2\varepsilon_0\frac{K(k_s)}{K(k_s')}.
\end{align}

Therefore, the total geometric inductance and capacitance per unit length of the CPW cross-section in flip-chip geometry, when the CPW faces a metal ground plane on the opposing chip, are given in (\ref{eqn:Lg_theory_topMetal}) and (\ref{eqn:C_theory_topMetal}) in the main text.

\section{Extraction of magnetic penetration depth}
\label{appendix: lambda}
To determine the magnetic penetration depth $\lambda_m$ of our superconducting films, we compare 3D-simulated and measured resonator frequencies (this time from single-chip devices, not flip-chip). We use $L_l^k$ as a fitting parameter to account for the discrepancy between measured and simulated frequencies, since the model uses 2D sheets with perfect-\textit{E} boundary condition, which does not include kinetic inductance. 

\begin{figure}[ht]
    \centering
    \includegraphics[width=\linewidth]{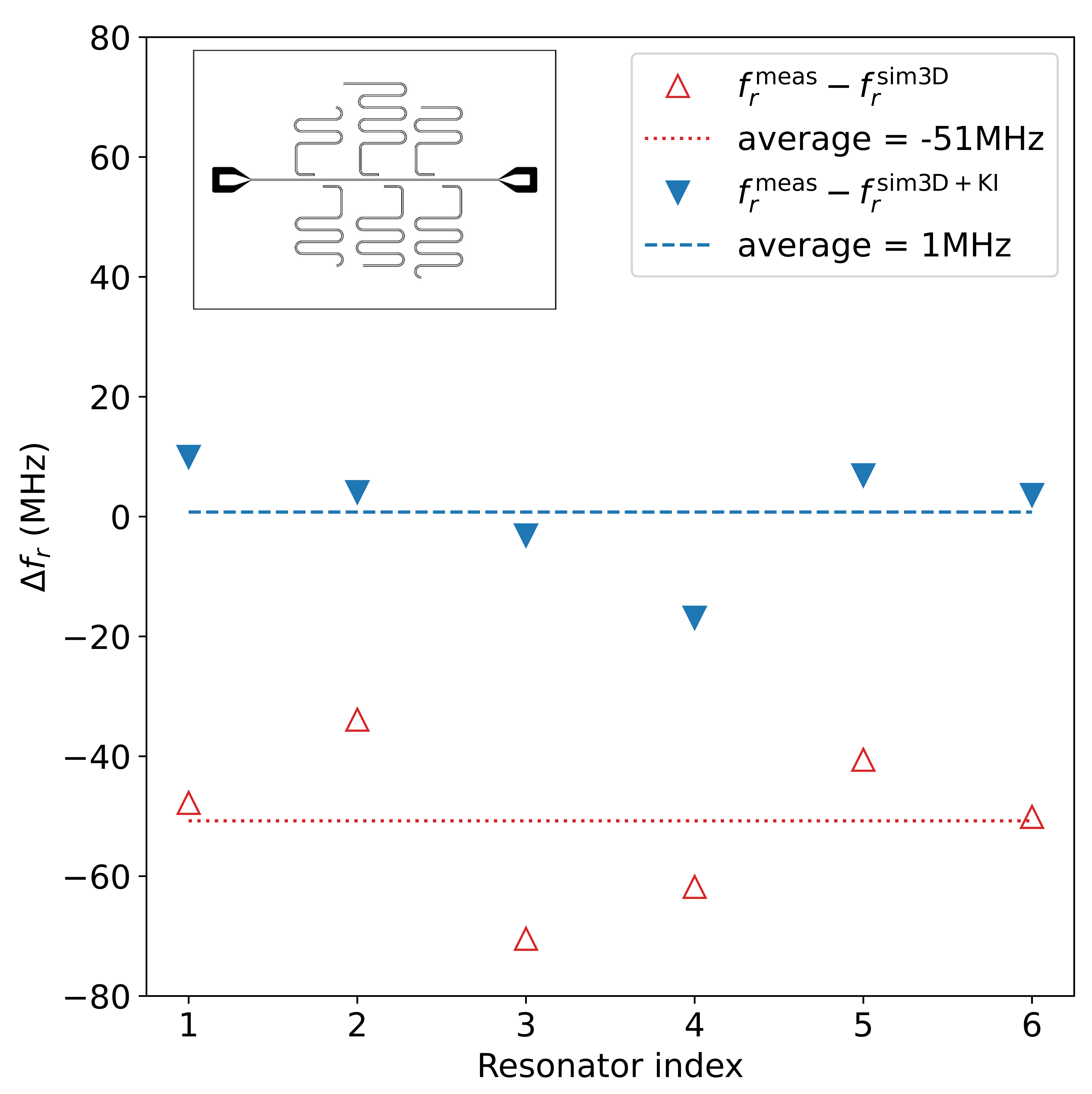}
    \caption{Difference between simulated and measured resonator frequencies. Each point represents the average of seven nominally identical resonators on different chips (single-chip, not flip-chip). The blue, downward-pointing triangles include the correction from kinetic inductance whereas the red, upward-pointing triangles do not. The inset shows the chip design with $w_r=20\,\upmu$m, $g_r=10\,\upmu$m.}
    \label{fig:ResFreq_Sim_vs_Meas_noTop}
\end{figure}

We measured six resonators on each of seven identical chips, see Fig.~\ref{fig:ResFreq_Sim_vs_Meas_noTop}. We simulated the supercurrent density $J_z$ in COMSOL~\cite{Niepce2020}. We write the permittivity of the film as
\begin{equation}
    \varepsilon_{metal} = \varepsilon_0 - \frac{1}{\omega^2\mu_0\lambda_m^2} ,
    \label{eqn: epsilon_metal}
\end{equation}
where $\omega = 2\pi f$ and $f=6.6$~GHz is the frequency of the alternating current. The imaginary term of $\varepsilon_{metal}$ is negligible.

We obtain $\lambda_m=83$~nm for our 150~nm thick aluminum film. This is somewhat larger than the literature value of $50\,$nm \cite{orlando1991,CharlesP1996}, suggesting that our film is in the dirty limit, with mean free path smaller than the film thickness~\cite{Miller1959,Dressel2013,Gubin2005}. 

\section{Effective length due to circular coupling structure}
\label{appendix: eff length}
\begin{figure}[ht]
    \centering
    \includegraphics[width=\linewidth]{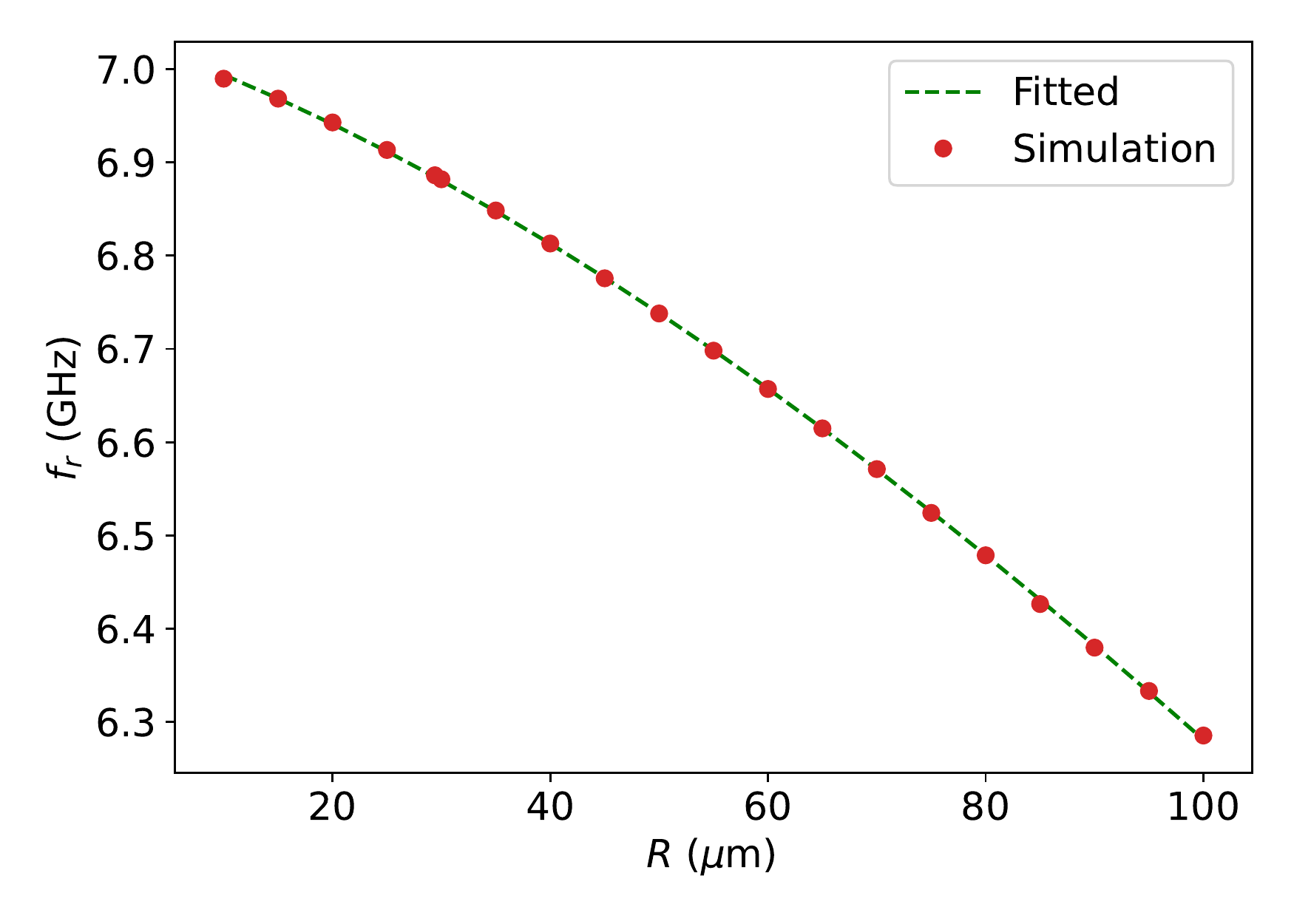}
    \caption{Decreasing resonator frequency $f_r$ due to the increase of the radius $R$, i.e., the increase in effective length of the circular coupling structure.}
    \label{fig:ResFreq_Leff_topMetal}
\end{figure}

In order to determine the effective length of the circular resonator-qubit coupling pad shown in Fig.~\ref{fig:3D2DRes_topMetal}(b), we compare the resonator frequency from 3D FEM simulation with 
\begin{equation}
    f_r = \frac{\beta}{l_r+\alpha_1R^2+\alpha_2R}.
    \label{eqn: fr_2D_effLen}
\end{equation}
Here $\beta=1/(4\sqrt{L_l^g C_l})$ is a fixed value for given $w_r$, $s_r$, $h_s$, $h_b$, and $\varepsilon_r$. 
Fig.~\ref{fig:ResFreq_Leff_topMetal} shows the simulated resonator frequency vs.\@ the radius $R$ of the coupling pad, for inter-chip spacing $h_s=8\,\upmu$m, and a fit to (\ref{eqn: fr_2D_effLen}). 

We obtain the fitted parameters $\alpha_1 = 0.032\,\upmu$m$^{-1}$ and $\alpha_2 = 2.9$ at $l_r = 5056\,\upmu$m. We use these fitted parameters to obtain the effective length at different $l_r$ and $h_s$ during $f_r$ and $Q_c$ calculations in the main text. Repeated simulations at different $l_r$ show that the change of calculated effective length ($\alpha_1R^2+\alpha_2R$) at a given $R$ is less than 3\%.

We also repeat the simulations at different $h_s$, from $1\,\upmu$m to $60\,\upmu$m, and find that for our maximum investigated $R=100\,\upmu$m, the change of the calculated effective length, is less than $113\,\upmu$m ($18\%$), compared to its value for $h_s=8\,\upmu$m. Smaller $R$ gives a smaller change, and we see that the change is kept at this value when $h_s\rightarrow\infty$. With $l_r=5056.4\,\upmu$m, the corresponding change of the resonator's frequency is less than $2\%$.

\section{Computational resources required by 2D and 3D FEM simulation}
\label{appendix: computing resource}
In this appendix we show the comparison of required computational resources between 2D and 3D FEM simulations.

We did 3D FEM simulations~\cite{ansys3D} of 25 resonators shown in Fig.~\ref{fig:ResFreq_Cal_2D_vs_Meas_topMetal}(a). We use the same setting as in \cite{Kosen2021} with \textit{Maximum Delta Frequency Per Pass} of 0.1\% for the Eigenmode solver, and \textit{Maximum Delta S} of 0.02 for the Driven Model solver. Across 25 resonators, the average spent CPU time was 32~h, and maximal memory allocation during Eigenmode simulation was 58~GB.

We conducted 2D FEM simulations~\cite{ansys2D} of the flip-chip CPW cross-section with chip spacing $h_s$ from 7 to $9~\upmu$m and step $0.2~\upmu$m. We use \textit{Percentage error} of 0.01\% for both parameter convergence criteria \textit{C only} and \textit{L only}. The average spent CPU time was 1.7~min, and maximal memory allocation was 64~MB. Compared to 3D FEM simulations, the CPU time and memory allocation were reduced by a factor of 1000 at each simulation.

In addition, we have also documented the used computational resources in COMSOL when simulating $J_z$ over the metal thin films, with $h_s$ from 7 to $9~\upmu$m with step $0.1~\upmu$m. The average solution time was 33~s and the allocated memory was 7.5~GB.

\section{Cross-sectional Q-factor resulting from lossy dielectric interfaces}
\label{appendix: participation ratio simulation}
In this appendix, we calculate the Q-factor, $Q_{pr}$, of the resonator's CPW line cross-section in flip-chip architecture, based on the simulated participation ratio of different domains~\cite{Wenner2011}. We use a similar 2D FEM simulation setting as that in~\cite{Kosen2021}, with relative permittivities $(4,4,7)$, loss tangents $(10^{-3},10^{-3},10^{-3})$, and film thicknesses $(2, 0.5, 2)\,$nm for the Substrate-Air, Substrate-Metal, and Metal-Air dielectric interfaces in the simulation. Other geometry parameters of the CPW line cross-section are the same as in Section~\ref{section: 2D cross-section calculations} when comparing conformal mapping techniques with 2D FEM simulation.

The Q-factor $Q_{pr}$ is calculated using
\begin{equation}
    1/Q_{pr} = \sum_ip_i\tan\delta_i,
    \label{eqn: fr_2D_effLen}
\end{equation}
where $p_i$ is the simulated participation ratio for each domain $i$ and $\tan\delta_i$ is the domain's loss tangent.

\begin{figure}[ht]
    \centering
    \includegraphics[width=\linewidth]{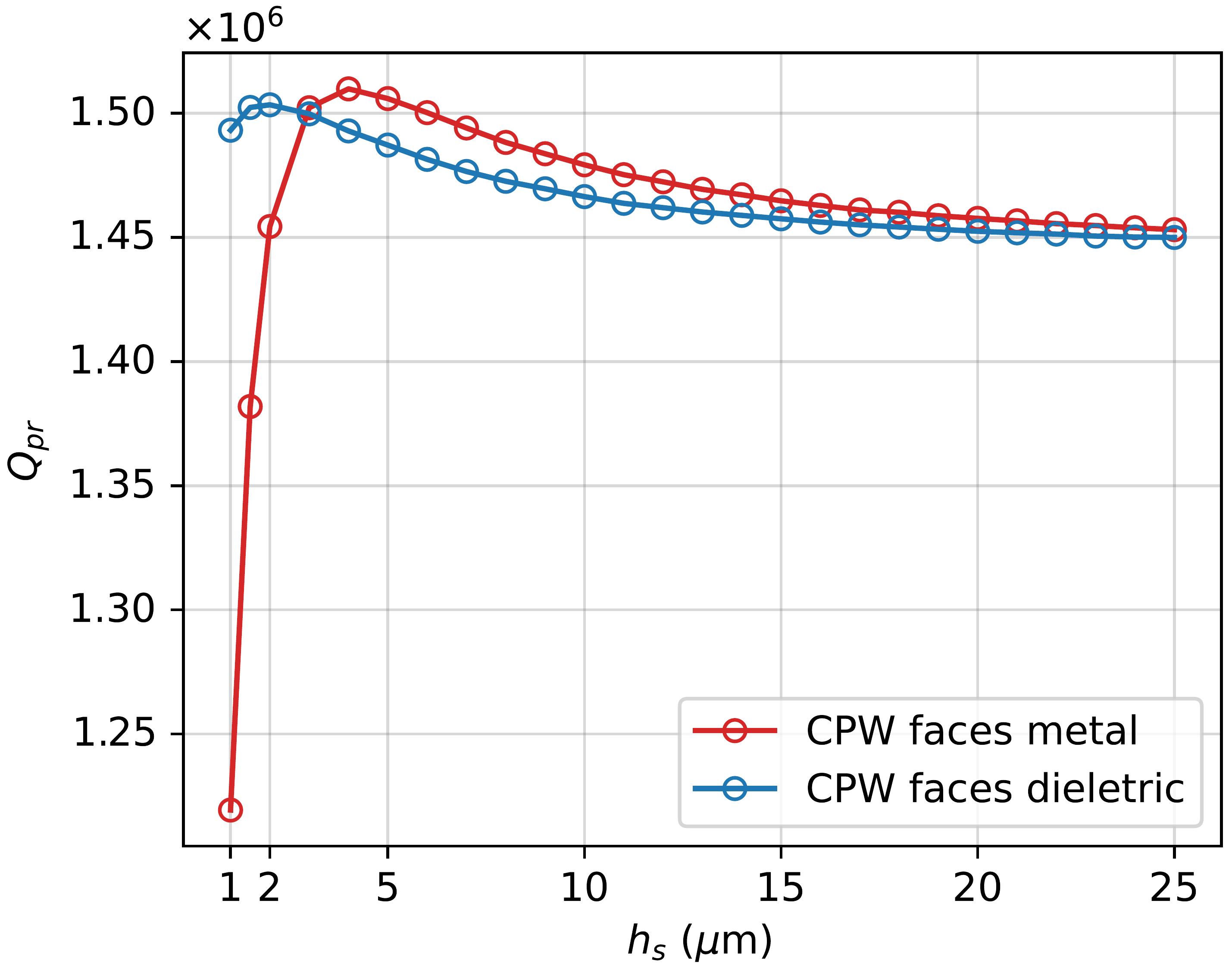}
    \caption{Cross-sectional Q-factor when the resonator's CPW line faces either the metal ground plane or the dielectric substrate of the qubit tier under different inter-chip spacings $h_s$.}
    \label{fig:Qfactor_2D}
\end{figure}

Fig.~\ref{fig:Qfactor_2D} shows the comparison of $Q_{pr}$ between the resonator's CPW line facing either the metal ground plane or the dielectric substrate of the qubit tier under different inter-chip spacings $h_s$. As $h_s$ decreases, there is a small increase of $Q_{pr}$ until $h_s$ is below a certain threshold. We also notice that $Q_{pr}$ is slightly lower when the CPW line is facing a dielectric substrate on the qubit tier. The Q-factor drops significantly in the case when the resonator CPW line is facing metal, which is caused by the increased electric field strength inside the Metal-Air dielectric interfaces when $h_s$ is very small~\cite{Kosen2021}.

\section*{Acknowledgment}
We thank Martí Gutierrez Latorre for his support on supercurrent density simulation in COMSOL. We are also grateful for discussions with David Niepce, Liangyu Chen, Janka Biznárová, Emil Rehnman, Christopher Warren, Anita Fadavi Roudsari and Per Delsing, and technical assistance with cryogenics from Alexey Zadorozhko and Liangyu Chen.

Author contributions: Hang-Xi Li conducted the analytical calculations, the numerical simulations and the writing of the paper. Daryoush Shiri provided critical inputs on the analytical calculations and supervised the work. Daryoush Shiri, Hang-Xi Li and Sandoko Kosen developed the idea of inter-chip-spacing insensitive flip-chip resonator design. Sandoko Kosen and Hang-Xi Li designed the multi-qubit flip-chip-integrated quantum processor. Marcus Rommel, Sandoko Kosen, Andreas Nylander, and Hang-Xi Li fabricated the two chips of the processor at Chalmers, followed by Marco Caputo, Kestutis Grigoras and Leif Grönberg conducting flip-chip bonding and gap measurements at VTT. Robert Rehammar, Giovanna Tancredi, and Sandoko Kosen provided the package of the processor and gave design advice on the processor's interface to the package. Hang-Xi Li and Sandoko Kosen measured the flip-chip resonators within the processor. Lert Chayanun fabricated and measured the single-chip resonators. Joonas Govenius supervised the work at VTT. Jonas Bylander supervised the whole work at Chalmers.

\ifCLASSOPTIONcaptionsoff
  \newpage
\fi

\bibliographystyle{IEEEtran_modified}
\bibliography{ref}

\EOD

\end{document}